# Two-dimensional phase diagram of the charge density wave in doped $CsV_3Sb_5$

Linwei Huai[1,#], Hongyu Li[1,#], Yulei Han[2,#], Yang Luo[1,#], Shuting Peng[1], Zhiyuan Wei[1], Jianchang Shen[1], Bingqian Wang[1], Yu Miao[1], Xiupeng Sun[1,3], Zhipeng Ou[1], Bo Liu[1], Xiaoxiao Yu[1], Ziji Xiang[1], Min-Quan Kuang[4], Zhenhua Qiao[1,*], Xianhui Chen[1,*] and Junfeng He[1,*]

[1]Department of Physics and CAS Key Laboratory of Strongly-coupled Quantum Matter Physics, University of Science and Technology of China, Hefei, Anhui 230026, China

[2]Department of Physics, Fuzhou University, Fuzhou, Fujian 350108, China

[3]Hefei National Laboratory, University of Science and Technology of China, Hefei 230088, China

[4]Chongqing Key Laboratory of Micro & Nano Structure Optoelectronics, and School of Physical Science and Technology, Southwest University, Chongqing 400715, China

[#]These authors contributed equally to this work.

*To whom correspondence should be addressed: Z.Q. (qiao@ustc.edu.cn), X.C. (chenxh@ustc.edu.cn), J.H. (jfhe@ustc.edu.cn).

Kagome superconductors $AV_3Sb_5$ (A = K, Rb and Cs) have attracted much recent attention due to the coexistence of multiple exotic orders. Among them, the charge density wave (CDW) order has been shown to host various unconventional behaviors. Here, we investigate the CDW order by a combination of both bulk and surface doping methods. While element substitutions in bulk doping change both carriers and the crystal lattice, the surface doping primarily tunes the carrier concentration. As such, our results reveal a two-dimensional phase diagram of the CDW in doped $CsV_3Sb_5$. In the lightly bulk doped regime, the existence of CDW order is reversible by tuning the carrier concentration. But excessive bulk doping permanently destroys the CDW, regardless of the carrier doping level. These results provide insights to the origin of the CDW from both electronic and structural degrees of freedom. They also open an avenue for manipulating the exotic CDW order in Kagome superconductors.

Kagome metals have attracted much attention due to the special lattice structure and associated physical properties [1-4]. The ongoing interest is further energized by Kagome superconductors $AV_3Sb_5$ (A = K, Rb and Cs), in which many exotic phenomena have been observed [5-50], ranging from superconductivity [5-9], charge density wave (CDW) [5,6,10-19], nematic order [20-21], pair density wave [22], topological states [5,23-24] and the time-reversal symmetry breaking state [17,25-27]. While the origins of these phenomena remain unclear, it is interesting that most of them are closely related to the CDW state [5,6,10-19]. In this regard, it is important to understand the driving mechanism of the CDW. Two scenarios have been primarily suggested [15,29-36]. The first is associated with the electronic instability, presumably driven by Fermi surface nesting [15,29-32], and the second is related to the structural instability via electron-phonon coupling [15,32-36]. In order to examine the origin, doping evolution of the CDW has been experimentally investigated by chemical element substitutions [38-45]. Nevertheless, both the carrier concentration and the crystal lattice have been changed in this process. Surface doping has also been performed by Cs surface deposition [46]. Comparing to the bulk element substitutions, the surface doping primarily induces carriers (electrons) to the surface layers of the samples [46]. Its application on a pristine $CsV_3Sb_5$ compound has led to a monotonic suppression of the CDW order [46]. Despite the continuous efforts, it remains challenging to differentiate the roles played by electron and lattice degrees of freedom [38-46].

In this paper, we investigate the evolution of CDW order in $CsV_3Sb_5$ via a combination of both bulk and surface doping. Angle-resolved photoemission spectroscopy (ARPES) measurements, which are sensitive to both doping methods [51], have been carried out to track the evolution of the CDW order in this material. First, Ti substitution of V is applied to the bulk crystal of $CsV_3Sb_5$, which induces hole doping and modifies the V Kagome net simultaneously. Continuous Cs surface deposition is then carried out on the $CsV_{3-x}Ti_xSb_5$ samples, which gradually induces electrons to compensate the holes doped by the Ti substitution. It is interesting that the CDW order is reversible as a function of carrier concentration in the lightly Ti doped regime. This is evidenced by the CDW gap, which disappears with Ti doping, but reappears with Cs surface deposition. However, excessive Ti bulk doping permanently destroys the CDW order, which becomes irreversible by tuning the carrier concentration. These results reveal a two-dimensional phase diagram of the CDW order in doped $CsV_3Sb_5$, and provide key insights to the associated driving mechanism.

Fig. 1 shows the electronic structure of $CsV_{3-x}Ti_xSb_5$ as a function of Ti bulk doping (see Supplementary Note 1 and Supplementary Figure 1 for the sample characterization), measured at a low temperature (10K). The photoemission measurements are carried out along *Γ-K-M* direction, where the CDW gap is most clearly revealed in the pristine $CsV_3Sb_5$ compound [32,47,48]. In order to visualize the evolution of the gap, low energy spectra around the van Hove singularity (vHs, marked by the red dotted boxes in Fig. 1a-d) are symmetrized with respect to the Fermi level ($E_F$) (Fig. 1e-h). It is clear that the CDW gap decreases with Ti doping and disappears at doping levels *x*≥0.13 (Fig. 1g,h). This is quantified by the corresponding energy distribution curves (EDCs) (Fig. 1i-p), where both Fermi-Dirac divided and symmetrized EDCs reveal the same doping evolution. Momentum dependence of the CDW gap is also examined on a Ti doped sample (*x*=0.03), which is largely consistent with that of the pristine $CsV_3Sb_5$ compound, but with an overall smaller magnitude of the gap (Fig. 1q-s).

Next, Cs surface doping is gradually carried out on the $CsV_{3-x}Ti_xSb_5$ (*x*=0.13) sample at a fixed low temperature (10K). As illustrated in Fig. 1 (also shown in Fig. 2a,e,f), the CDW gap is completely suppressed by Ti doping in this sample. Upon Cs surface doping, a slight downward shift of the V bands is observed (Fig. 2a-d, for example, see the band top near K point, marked by the black triangles). Most strikingly, a spectral weight suppression starts to appear near $E_F$, indicating the possible reappearance of an energy gap. This behavior is quantitatively revealed by the EDCs around the vHs region, where an energy gap gradually forms at $E_F$ as a function of the Cs surface doping (Fig. 2e-l). In order to examine whether the gap opening is associated with the reappearance of the CDW order or created by extrinsic effects (e.g. a possible localization effect induced by disorders), the Cs surface doping is repeated on the $CsV_{3-x}Ti_xSb_5$ (*x*=0.13) sample, but at a much higher temperature (100 K). Different from the behavior at low temperature, now the system remains gapless after the Cs surface doping (Fig. 2m-r). Nevertheless, the energy gap starts to appear on the band near the vHs (with V d-orbitals) when the temperature is cooled down to ~40 K (Fig. 2o,p). This gap becomes more evident at a lower temperature of 10 K (Fig. 2o,p). On the contrary, the band around the Γ point (with the Sb P-orbital) remains gapless at all temperatures (Fig. 2q,r). The above temperature and momentum (orbital) dependent behaviors of the gap are highly consistent with those of the CDW gap, demonstrating the reappearance of the CDW order in this system.

In order to further understand the evolution of the CDW order, the surface Cs doping is carried out on another CsV$_{3-x}$Ti$_x$Sb$_5$ sample, but with a higher bulk doping level ($x$=0.39) at 10 K. Apparently, the CDW order is completely suppressed in this sample before the Cs surface doping, evidenced by the absence of the CDW gap (Fig. 3a,e,i,j). Upon the Cs doping, an overall downward shift of the V bands is observed as before (Fig. 3a-d). We note that excessive Cs doping has been applied on this sample (Fig. 3c,d) to ensure that the holes induced by the higher level of Ti substitution are sufficiently compensated and a similar total carrier concentration is achieved as before. This is evidenced by similar positions of the energy features on the band structure of the two samples after Cs deposition (e.g. compare the band top between Γ and K in Fig. 2c,d, n and Fig. 3c,d, marked by the black triangles). However, distinct from the earlier case, no evidence of gap opening is observed on the CsV$_{3-x}$Ti$_x$Sb$_5$ ($x$=0.39) sample with Cs surface doping. This gapless state remains robust, regardless of the electron doping level induced by the Cs surface deposition (Fig. 3e-p).

The above bulk and surface doping dependent measurements (Fig. 4a-c) reveal a two-dimensional phase diagram of the CDW in doped CsV$_3$Sb$_5$ (Fig. 4d). First, the CDW order is monotonically suppressed as a function of the Ti bulk doping, and the CDW gap disappears at the doping level of $x$=0.13. Then, continuous surface doping at the fixed Ti doping level ($x$=0.13) leads to the reappearance of the CDW gap. However, the Cs surface doping shows little effect on the sample with a higher Ti doping level of $x$=0.39. The sample remains gapless, regardless of the surface doping level.

Finally, we discuss the implications of such an experimental phase diagram. First, the Ti bulk doping changes both the carrier concentration and the lattice of the material. On one hand, the substitution of V atoms by Ti atoms effectively induces holes into the Kagome layer. On the other hand, the new Ti atoms would inevitably modify the original V Kagome net and affect the lattice vibration. The totality of these effects results in the suppression of the CDW order. Second, the Cs surface doping primarily induces electrons to the sample. In pristine $CsV_3Sb_5$, a suppression of CDW with Cs surface doping has been reported [46], where the surface induced electrons may shift the carrier concentration away from the optimal doping. On the other hand, the electrons induced by Cs surface deposition would compensate the holes doped by Ti substitution in the $CsV_{3-x}Ti_xSb_5$ compound. The reappearance of the CDW order in the lightly Ti doped regime ($x$=0.13) demonstrates that the carrier concentration is important for the CDW. It also provides a tuning knob to manipulate the CDW order. However, the suppression of CDW order becomes irreversible by surface doping in the heavily bulk doped regime ($x$=0.39), which indicates that the carrier concentration is not the only controlling parameter of the CDW. In order to better understand the role of lattice, we have calculated the total energy profiles of $CsV_{3-x}Ti_xSb_5$ [15,52,53] at the Ti doping levels of $x$=0 (pristine $CsV_3Sb_5$, Fig. 4e), $x$=0.125 (Fig. 4f) and $x$=0.375 (Fig. 4g) (see Supplementary Note 2 for the details of the calculation), which are very close to the doping levels of the samples measured in the experiments. It is clear that the Kagome structure is unstable in the pristine compound, in which the Inverse Star of David (ISD) structure has the lowest total energy [15] (see Supplementary Figure 2 for the schematics of Kagome, Star of David and Inverse Star of David structures). This lattice instability persists to the Ti doping level of $x$=0.125 (Fig. 4f), but disappears at $x$=0.375 (Fig. 4g). These results, when combined with the experimental observations, point to a unified understanding of the two-dimensional phase diagram. The coexistence of lattice instability and appropriate amount of carrier concentration is needed to establish the CDW order in the $CsV_3Sb_5$ system. In this regard, the lattice instability naturally provides a tendency towards the CDW transition, and the carrier concentration serves as a tuning knob to change the magnitude (and possibly correlation length) of the CDW order. When the lattice instability persists (for example, in a finite Ti doping range around x=0.13), the CDW order is reversible by tuning the carrier concentration. In addition, the change of carrier concentration may also affect the electronic correlation in the material system (for example, via screening effect). In this sense, our observations can also reconcile the earlier reports that multiple unconventional properties in the CDW state are associated with electronic correlation [20,22,25-27], but the electronic instability itself is insufficient to drive the CDW phase transition [34,35,49,54].

In summary, by utilizing high-resolution ARPES measurements, we have revealed a two-dimensional phase diagram of the CDW in CsV$_3$Sb$_5$ as a function of both Ti bulk substitution and Cs surface deposition. The distinct evolutions in this phase diagram reveal the roles played by both electrons and lattice. These observations provide key insights to understand the driving mechanism of the CDW order in Kagome metals.

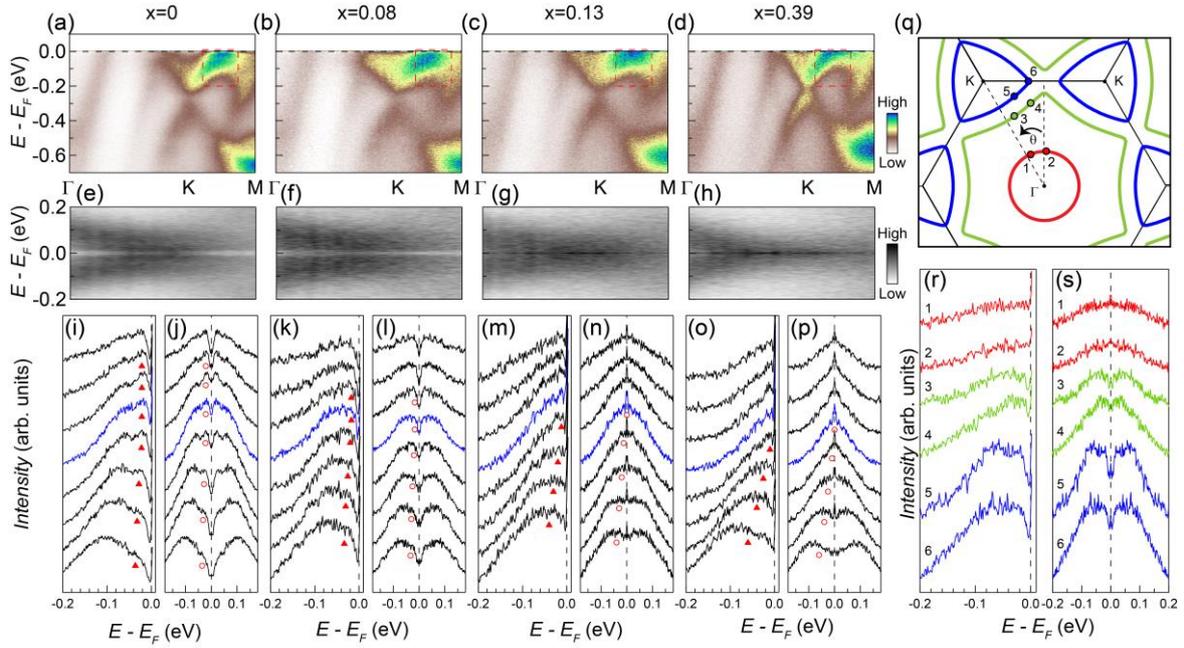

**Fig.1 Electronic structure of $CsV_{3-x}Ti_xSb_5$ as a function of the Ti substitution level, measured at 10K.** a-d Photoelectron intensity plot of the band structure along the Γ-K-M direction on $CsV_{3-x}Ti_xSb_5$ samples with x=0(a), 0.08(b), 0.13(c) and 0.39(d). e-h symmetrized low energy spectrum near the vHs region in a-d. i-p EDCs in the red dotted boxes in a-d. The EDCs are divided by the Fermi-Dirac function (i, k, m, o) and symmetrized with respect to $E_F$ (j, l, n, p), respectively. EDCs at the Fermi momenta are highlighted in blue. The EDC peaks are marked by triangles and circles. q Schematic Fermi Surface of $CsV_{3-x}Ti_xSb_5$ (x=0.03). r-s EDCs at representative momentum points. The momentum locations of the points are shown in q.

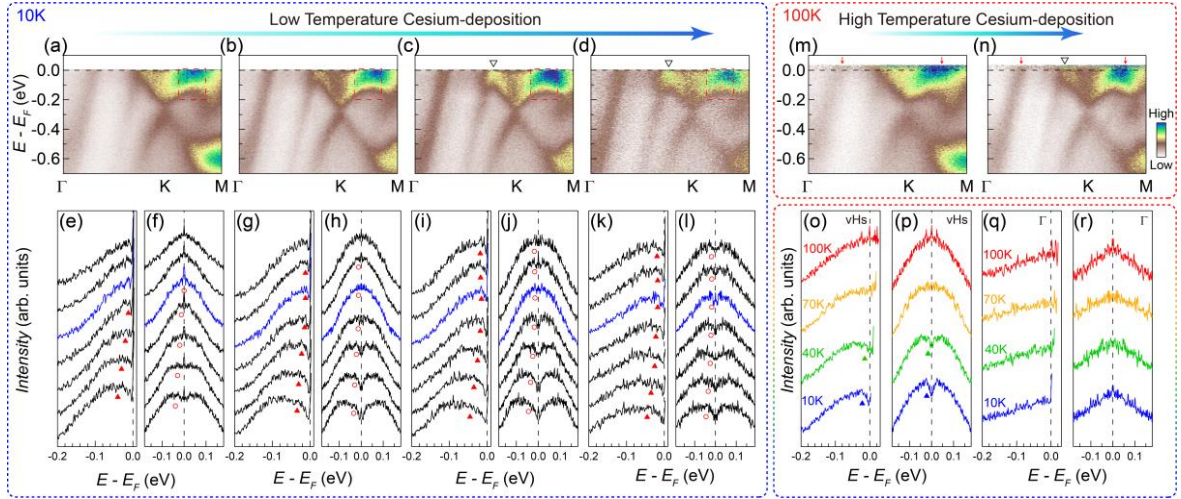

***Fig. 2 Evolution of the electronic structure with Cs surface doping on the CsV$_{3-x}$Ti$_x$Sb$_5$ (x=0.13) sample.*** *a-d Photoelectron intensity plot of the band structure along the Γ-K-M direction as a function of Cs surface doping at 10K. e-l EDCs in the red dotted boxes in a-d. The EDCs are divided by the Fermi-Dirac function (e, g, i, k) and symmetrized with respect to $E_F$ (f, h, j, l), respectively. EDCs at the Fermi momenta are highlighted in blue. m-n Photoelectron intensity plot of the band structure along the Γ-K-M direction before(m) and after(n) Cs surface deposition at 100K. The red arrows mark the Fermi momenta. o-p Temperature dependence of the EDC at the Fermi momentum near the vHs. To reveal the gap opening, the EDC is divided by the Fermi-Dirac function (o) and symmetrized with respect to $E_F$ (p), respectively. q-r, Same as o-p, but measured at the Fermi momentum near Γ. The black triangles in c, d and n mark the band top between Γ and K.*

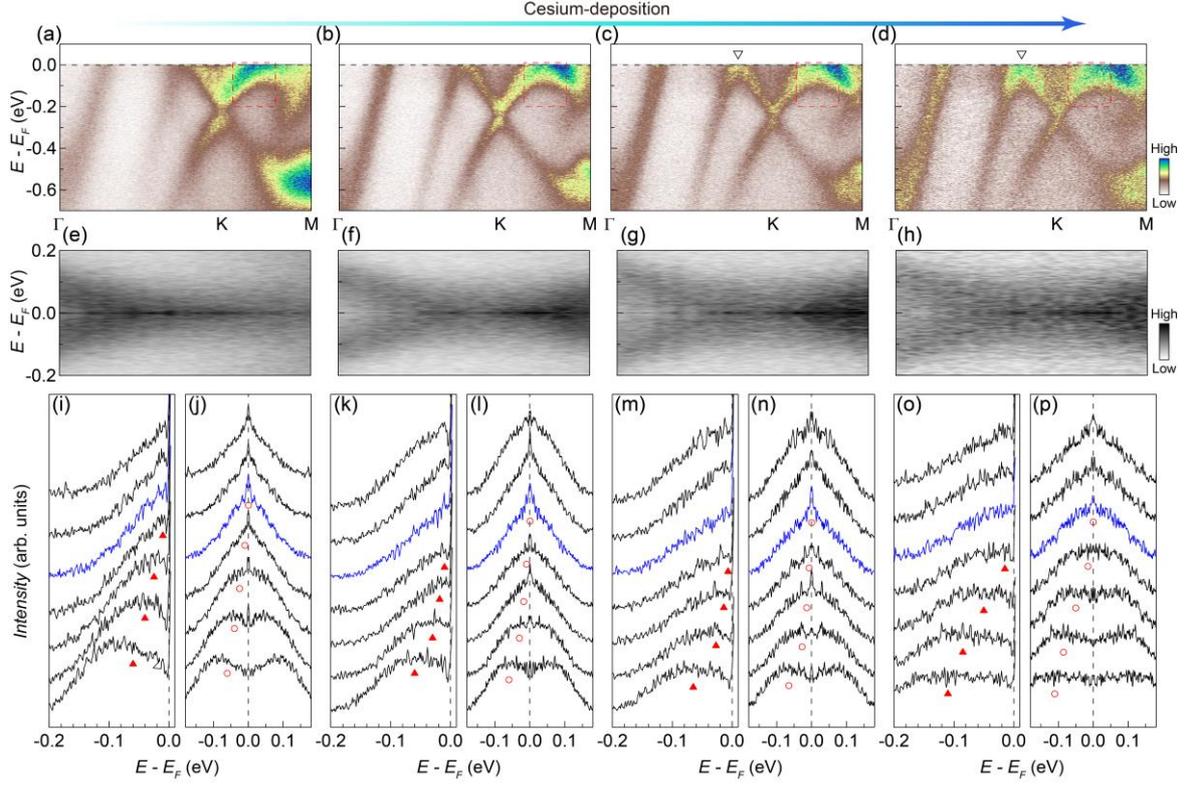

***Fig. 3 Evolution of the electronic structure with Cs surface doping on the $CsV_{3-x}Ti_xSb_5$ (x=0.39) sample.*** *a-d Photoelectron intensity plot of the band structure along the Γ-K-M direction as a function of Cs surface doping at 10K. The black triangles in c and d mark the band top between Γ and K. e-h symmetrized low energy spectrum near the vHs region in a-d. i-p EDCs in the red dotted boxes in a-d. The EDCs are divided by the Fermi-Dirac function (i, k, m, o) and symmetrized with respect to $E_F$ (j, l, n, p), respectively. EDCs at the Fermi momenta are highlighted in blue.*

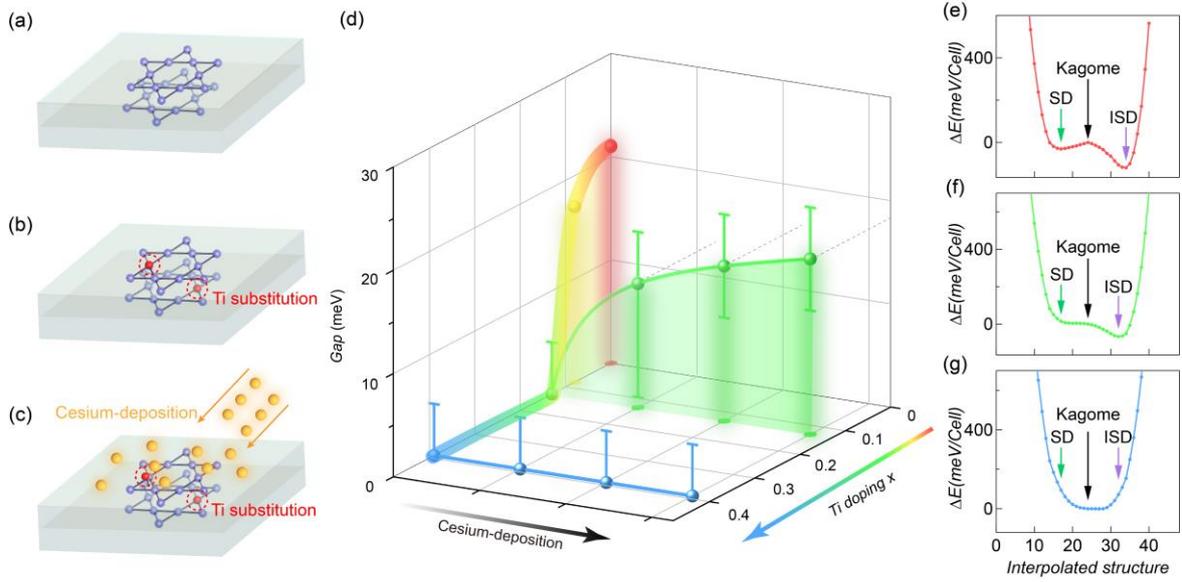

*Fig. 4 Two-dimensional phase diagram of the CDW in doped CsV$_3$Sb$_5$ and total energy profiles of CsV$_{3-x}$Ti$_x$Sb$_5$.* a-c Schematics of the doping processes. d two-dimensional phase diagram of the CDW gap in CsV$_3$Sb$_5$ as a function of Ti substitution and Cs surface deposition. e-g Calculated total energy of pristine CsV$_3$Sb$_5$ (e), CsV$_{3-x}$Ti$_x$Sb$_5$ (x=0.125) (f), and CsV$_{3-x}$Ti$_x$Sb$_5$ (x=0.375) (g). ΔE represents the relative total energy with respect to that of the Kagome structure. It is shown in the 2×2×2 supercell (72 atoms) for all cases.